\documentstyle[epsfig,12pt,preprint,tighten,aps]{revtex}
\begin{document}

\draft

\title{\rightline{{\tt March 1999}}
\rightline{{\tt UM-P-99/05}}
\rightline{{\tt RCHEP-99/01}}
\ \\
Implications of TeV scale $SU(4) \otimes SU(2)_L \otimes
SU(2)_R$ quark-lepton unification}
\author{R. Foot\footnote{foot@physics.unimelb.edu.au} and G. Filewood}
\address{
School of Physics\\
Research Centre for High Energy Physics\\
The University of Melbourne\\
Parkville 3052 Australia }

\maketitle

\begin{abstract}
The alternative $SU(4) \otimes SU(2)_L \otimes SU(2)_R$
gauge model, which allows unification of the quarks and leptons
at the TeV scale, is studied in detail. 
We discuss the implications for nucleon decay, B and K rare
meson decays and neutrino masses.
We also explain how this model solves the gauge
hierarchy problem {\it without} using supersymmetry or
extra large dimensions.

\end{abstract}

\newpage
\section{Introduction}
\vskip 0.5cm

There has been some discussion lately about the gauge hierarchy
problem\cite{lar}. We would like to contribute to this discussion by
suggesting the following rather simple solution:
If the scale of new physics is $M_{new}$ then a gauge hierarchy can be
avoided provided that $M_{new}$ is 
not too different from the weak scale, $M_{weak}$.
This suggests the following
rough upper limit
\begin{equation}
M_{new} \stackrel{<}{\sim}\ few \ TeV 
\label{1}
\end{equation}
We prefer to exclude gravity from our discussion
for the obvious reason that it is not a 
well understood quantum theory. Despite the large
value of $M_P \sim 10^{19} \ GeV$ it is not clear
whether this poses a fine-tuning problem or not.
The mere existence of the two disparate scales
$M_{weak}$ and $M_P$ does not {\it necessarily} imply a fine-tuning
problem, just like the existence of the disparate scales
$\Lambda_{QCD}$ and $M_{weak}$ does not imply a fine tuning problem
in the standard model. Thus, we argue that so long
as $M_{new} \stackrel{<}{\sim} \ few \ TeV$ the gauge 
hierarchy problem can be avoided. Of course it 
should also be emphasised that the condition, Eq.(\ref{1}) is
of great practical importance
since it means that the theory can be subject to 
many experimental tests (in principle). 

Given the rather stringent requirement, Eq.(\ref{1}), one might
imagine that there is no new physics beyond the standard model. We
argue that this is unlikely for at least three reasons:
\begin{enumerate}
\item There is experimental and theoretical evidence for
neutrino masses. The experimental evidence comes from the
neutrino physics anomalies (such as
the atmospheric, solar, LSND anomalies), while the theoretical 
evidence comes from the electric charge quantization problem of the
minimal standard model\cite{cq};

\item Each generation contains five distinct fermionic gauge multiplets;

\item The standard model is a bit ugly because it contains $20$
theoretically unconstrained parameters.
\end{enumerate}

Let us first remark that the model to be discussed in this
paper has many more parameters than the standard model,
so that we have certainly not made any progress on the parameter
problem. However the model does partially address the
other two points identified above.
One of the reasons that each generation contains five distinct 
fermion multiplets is that the quarks and the leptons are
similar but lack any real symmetry in the standard model.
Thus one obvious way to improve on this is 
to embed the standard model into a gauge model with a symmetry
between the quarks and the leptons. Given the 
constraint, Eq.(\ref{1}) there are only two
possibilities that we are aware of. The first TeV scale quark-lepton
unified model was proposed in Ref.\cite{fl} where a leptonic 
$SU(3)_{\ell}$ colour group was assumed so that
a discrete $Z_2$ quark-lepton symmetry can be defined (the
$SU(3)_{\ell}$ gauge symmetry is assumed to be spontaneously 
broken at the TeV scale).
More recently, one of us\cite{foot} has also shown that it is possible
to modify the usual Pati-Salam model\cite{ps} 
such that the quarks and the leptons can be unified with
gauge group $SU(4) \otimes SU(2)_L \otimes SU(2)_R$ at the low
scale of about a TeV\footnote{
For some discussions of the usual Pati-Salam model
with a low symmetry breaking scale, see Ref.\cite{will}.
Note however that in the usual Pati-Salam model the lowest
possible value for the symmetry breaking scale is
still quite high. According to Ref.\cite{will2} it is
$m_{W'} \stackrel{>}{\sim} 13\ TeV$.}.
The purpose of this paper is to provide a systematic study of this 
alternative $SU(4) \otimes SU(2)_L \otimes SU(2)_R$ model
(or 422 model for short).

The outline of this paper is as follows: In section II we review
the basic structure of the alternative $SU(4) \otimes SU(2)_L
\otimes SU(2)_R$ model. In section III we investigate nucleon
decay in this model. As already noted in Ref.\cite{foot}, 
gauge interactions conserve a global baryon number, however
this symmetry can be broken by scalar interactions
in the Higgs potential. We show that the effect of the scalar 
mediated nucleon decay is to induce neutron decay $N \to \ell^+ \ell^- \nu$
(where $\ell = e, \mu$).
We provide a rough estimate of this decay rate which we
show is consistent with a TeV symmetry breaking scale.
In section IV we discuss rare $B, K$ meson decays.
These decays provide the main experimental bound on the model.
In section V we discuss neutrino masses in the model which
are naturally small, despite the TeV symmetry breaking scale.
In section VI we conclude.

\section{
The alternative $422$ model}

In this section, we review the alternative $422$ model.
For more details see
Ref.\cite{foot}.  The gauge symmetry of the model is 
\begin{equation}
SU(4) \otimes SU(2)_L \otimes SU(2)_R.
\label{uuuuu}
\end{equation}
Under this gauge symmetry the fermions of each generation transform 
in the anomaly free representations:
\begin{equation}
Q_L \sim (4,2,1),\  Q_R \sim (4, 1, 2), \ f_L \sim (1,2,2).
\label{2}
\end{equation}
The minimal choice of scalar multiplets which can both
break the gauge symmetry correctly and give all of the 
charged fermions mass is
\begin{equation}
\chi_L \sim (4, 2, 1), \ \chi_R \sim (4, 1, 2),\ \phi \sim (1,2,2).
\label{3}
\end{equation}
These scalars couple to the fermions as follows\footnote{
Note that we do not include a bare mass term
$m_{bare}\bar f_L (f_L)^c$, although such a term
is allowed by the gauge symmetry of the model.
We assume that $m_{bare} \ll M_{weak}$ so that it
can be neglected. This is not unreasonable, since the
bare mass scale is completely independent of the
weak scale.}:
\begin{equation}
{\cal L} = \lambda_1 \bar Q_L (f_L)^c \tau_2 \chi_R 
+ \lambda_2 \bar Q_R f_L \tau_2 \chi_L 
+ \lambda_3 \bar Q_L \phi \tau_2 Q_R  + 
\lambda_4 \bar Q_L \phi^c \tau_2 Q_R  
+ H.c.,
\label{4}
\end{equation}
where the generation index has been suppressed and $\phi^c = \tau_2 
\phi^* \tau_2$.  Under the $SU(3)_c \otimes U(1)_T$ 
subgroup of $SU(4)$, the $4$ representation has the branching rule,
$4 = 3(1/3) + 1(-1)$.  We will assume that the 
$T=-1, I_{3R} = 1/2 \ (I_{3L}=1/2)$ components of $\chi_{R} (\chi_L)$
gain non-zero Vacuum Expectation Values (VEVs) as well as 
the $I_{3L} = -I_{3R} = -1/2$ and $I_{3L} = -I_{3R} = 1/2$
components of the $\phi$.
We denote these VEVs by $w_{R,L}, u_{1,2}$ respectively.
In other words,
\begin{eqnarray}
\langle \chi_R (T = -1, I_{3R} = 1/2) \rangle = w_R, \
\langle \chi_L (T = -1, I_{3L} = 1/2) \rangle = w_L, 
\nonumber \\
\langle \phi (I_{3L} = -I_{3R} = -1/2)\rangle = u_1,\
\langle \phi (I_{3L} = -I_{3R} = 1/2)\rangle = u_2.
\end{eqnarray}
We will assume that the VEVs satisfy
$w_R > u_{1,2}, w_L$
so that the symmetry is broken as follows:
\begin{eqnarray}
&SU(4)\otimes  SU(2)_L \otimes SU(2)_R&
 \nonumber \\
&\downarrow \langle \chi_R \rangle&
\nonumber \\
&SU(3)_c \otimes SU(2)_L \otimes U(1)_Y &
\nonumber \\ 
&\downarrow \langle \phi \rangle, \langle \chi_L \rangle
\nonumber \\
&SU(3)_c \otimes U(1)_Q&
\end{eqnarray}
where $Y = T +2I_{3R}$ is the linear combination of 
$T$ and $I_{3R}$ which annihilates $\langle \chi_R \rangle$ 
(i.e. $Y\langle \chi_R \rangle = 0$)\footnote{
In Ref.\cite{foot} a distinct but physically equivalent
convention was used, which leads to $Y = T - 2I_{3R}$. 
The difference is just a $SU(2)_R$ rotation.}.  Observe that in 
the limit where $w_R \gg w_L, u_1, u_2$, the model 
reduces to the standard model. The VEV $w_R$ breaks the 
gauge symmetry to the standard model subgroup. This VEV 
also gives large $SU(2)_L \otimes U(1)_Y$ invariant masses to 
an $SU(2)_L$ doublet of exotic fermions, which  
have electric charges $-1, 0$.  We will denote these 
exotic fermions with the notation $E^-, E^0$.
These exotic fermions must have masses greater than about
$m_Z/2$ otherwise they would contribute to the $Z$ width.
Observe that the right-handed chiral components of 
the usual charged leptons are contained in $Q_R$. 
They are the $T=-1, I_{3R} = -1/2$ components.
The usual left-handed leptons are contained in 
$f_L$ along with the right-handed components 
(CP conjugated) of $E^0, E^-$.
It is instructive to write out the fermion multiplets
explicitly. For the first generation,
\begin{eqnarray}
Q_L = \left(\begin{array}{cc}
d_y & u_y   \\
d_g & u_g  \\
d_b & u_b  \\
E^- & E^0 
\end{array}
\right)_L,\
Q_R = \left(\begin{array}{cc}
d_y & u_y  \\
d_g & u_g \\
d_b & u_b \\
e & \nu  
\end{array}
\right)_R, \
f_L = \left(\begin{array}{cc}
\nu_L & (E^-_R)^c  \\
e_L & (E^0_R)^c 
\end{array}
\right)_L,
\end{eqnarray}
and similarly for the second and third generations.
In the above matrices the first column of $Q_L$ $(f_L,\ Q_R)$
is the $I_{3L} (I_{3R}) = -1/2$ component while the second
column is the $I_{3L} (I_{3R}) = 1/2$ component.
The four rows of $Q_L, Q_R$ are the four colours and
the rows of $f_L$ are the $I_{3L} = \pm 1/2$ components.
Observe that the VEVs $w_L, u_{1,2}$ have the quantum 
numbers $I_{3L} = -1/2, Y = 1$
(or equivalently $I_{3L} = 1/2, Y = -1$).
This means that the standard model subgroup, 
$SU(3)_c \otimes SU(2)_L \otimes U(1)_Y$
is broken to $SU(3)_c \otimes U(1)_Q$ in the usual way (with
$Q = I_{3L} + Y/2 = I_{3L} + I_{3R} + T/2$). 

\section{Scalar ${\chi}_{R}$ and  ${\chi}_{L}$
mediated baryon-number violating interactions}

One of the main constraints on unified
models is the empirical limit on nucleon decay. 
Baryon charge in the alternative
$422$ model is defined as
$B=(B'+T)/4$ where the $B'$ charges of 
$Q_L,Q_R,\chi_{L,R}$ are all
+1 and the $B'$ charges of $f_L,\phi$
are 0. (The $B'$ charges of the gauge bosons are also 0).
This baryon charge is conserved
by the gauge interactions and Yukawa Lagrangian,
Eq.(\ref{4}).
(While $B'$ and $T$ are both broken by the vacuum,
the combination $B'+T$ is unbroken, since
$(B'+T)\langle \chi_R \rangle = 
(B' +T)\langle \chi_L \rangle = 
(B' +T)\langle \phi \rangle = 0$)\footnote{
Note that the baryon charge [$B=(B'+T)/4$] of the quarks is $1/3$ and
the baryon charge of the leptons is $0$. Also it is 
straightforward to check that the 
baryon charge of the $SU(3)_c \otimes SU(2)_L \otimes U(1)_Y$
gauge bosons are also zero.}. 
Thus, the only part of the Lagrangian which can
potentially mediate nucleon decay is the Higgs
potential. It is quite easy to see that the only
gauge invariant and renormalizable terms in the Higgs
potential which break the $U(1)_B$ symmetry are
\begin{equation}
V_1 = \stackrel{\sim}{\lambda_1}
\chi_L^{-{1 \over 2}} \chi_L^{+{1 \over 2}} 
\chi_R^{-{1 \over 2}} \chi_R^{+{1 \over 2}} 
+ \stackrel{\sim}{\lambda_2}
\chi_L^{-{1 \over 2}} \chi_L^{+{1 \over 2}} 
\chi_L^{-{1 \over 2}} \chi_L^{+{1 \over 2}} 
+ \stackrel{\sim}{\lambda_3}
\chi_R^{-{1 \over 2}} \chi_R^{+{1 \over 2}} 
\chi_R^{-{1 \over 2}} \chi_R^{+{1 \over 2}} + H.c.,
\end{equation}
where the $I_{3L,R}$ quantum numbers have been explicitly shown
as superscripts. From Eq.(\ref{4}) the colour triplet components
of the $\chi's$ interact with the fermions as follows,
\begin{eqnarray}
{\cal L}^{\chi} &=& 
\lambda_1 \chi_R^{-{1 \over 2}} 
\left( \bar d_L(\nu_L)^c - \bar u_L (e_L)^c\right)
+ \lambda_1 \chi_R^{+{1 \over 2}} \left( 
\bar d_LE_R^-  - \bar u_L E_R^0\right)
\nonumber \\
&+& \lambda_2 \chi_L^{-{1 \over 2}}\left(
\bar d_R \nu_L + \bar u_R (E^-_R)^c\right)
- \lambda_2 \chi_L^{+{1 \over 2}}\left(
\bar d_R e_L + \bar u_R (E^0_R)^c\right) + H.c.
\end{eqnarray}
Thus, the Higgs potential term which leads to the
most significant nucleon decay is expected to be\footnote{
Observe that the term $  
 \chi_L^{-{1 \over 2}}
\chi_L^{+{1 \over 2}} \chi_L^{-{1 \over 2}}
\langle \chi_L^{+{1 \over 2}} \rangle
$ primarily mediates $\Delta^-$ (ddd) decay, while
$\chi_R^{-{1 \over 2}}
\chi_R^{+{1 \over 2}} \chi_L^{-{1 \over 2}}
\langle \chi_L^{+{1 \over 2}} \rangle,
\ \chi_R^{-{1 \over 2}}
\chi_R^{+{1 \over 2}} \chi_R^{-{1 \over 2}}
\langle \chi_R^{+{1 \over 2}} \rangle
$ terms can mediate nucleon decay but are suppressed because
the $\chi_R^{+{1 \over 2}}$ state couples to the weak - eigenstate $E$
(which contains only a tiny admixture of the light $e, \mu$
mass eigenstates).}
\begin{equation}
V_2 = \stackrel{\sim}{\lambda_1} \chi_L^{-{1 \over 2}}
\chi_L^{+{1 \over 2}} \chi_R^{-{1 \over 2}}
\langle \chi_R^{+{1 \over 2}} \rangle
+ H.c.
= \stackrel{\sim}{\lambda_1} w_R \epsilon^{ijk}
\chi_{Li}^{-{1 \over 2}}
\chi_{Lj}^{+{1 \over 2}}
\chi_{Rk}^{-{1 \over 2}}
+ H.c.
\label{uu}
\end{equation}
where we have made the $SU(3)_c$ indices explicit 
[$(i,j,k) \ \epsilon \ {1,2,3}$].
This term mediates neutron decay via the Feynman diagram
in Figure 1.

The matrix element will contain a term for the propagator
of each scalar $\chi$ which will contribute a factor 
of ${{m_{\chi}^{-2}}}$. Thus the matrix element
will be proportional to $m_{\chi_R}^{-2}m_{\chi_L}^{-4}$
and the decay time, being proportional to the inverse
square of the matrix element will be of the form;
\begin{equation}
{\large \tau}_N \sim
\left( {4\pi \over \stackrel{\sim}{\lambda}_1 w_R}\right)^{2}
\lambda_1^{-2} \lambda_2^{-4} m\chi_R^{4}m\chi_L^{8} m_N^{-11},
\end{equation}
where $\lambda_1$ and $\lambda_2$ are the 
dimensionless coupling constants from the
interaction Lagrangian Eq.(\ref{4}), $\stackrel{\sim}{\lambda}_1 w_R$
is from the trilinear scalar interaction term, Eq.(\ref{uu}).
The neutron mass $m_N$ has been introduced as a 
dimensional factor because we are studying
neutron decay. Observe that the $\lambda_2$ Yukawa coupling
is proportional to the electron mass, so that
$\lambda_2 = m_e/w_L$. Strictly, the only information
that we know about $w_L$ is that
$u_1^2 + u_2^2 + w_L^2 \simeq (250 GeV)^2$, so 
that the most natural value for
$\lambda_2$ is $\lambda_2 \sim 10^{-5}$.
Thus, with this in mind, we have
\begin{equation}
{\large \tau}_N \sim {1 \over \lambda_1^2 \stackrel{\sim}{\lambda}_1^2}
\left( {10^{-5} \over \lambda_2}\right)^4
\left({TeV \over w_R}\right)^2
\left( {m_{\chi_R} \over TeV}\right)^4
\left( {m_{\chi_L} \over TeV}\right)^8
10^{21} \ years.
\end{equation}
The current experimental bound on the (bound)
neutron decay mode $N \to e \bar e \nu$ is
${\large \tau}_N \stackrel{>}{\sim} 7 \times 10^{31}\ years$ at 
90\% C.L.\cite{exp}.
This bound suggests $\lambda_1 \stackrel{\sim}{\lambda}_1 
\stackrel{<}{\sim} 10^{-5}$, which is not a very stringent limit.
Thus, clearly this model is not significantly constrained
by current limits on nucleon decay.
Obviously, if a $N \to e\bar e \nu$ signal were to be
experimentally observed, then this would be compatible with
this model.  Finally, note that we have implicitly assumed that
the scalars $\chi_L, \chi_R$ coupled the first generation
quarks, $u,d$ with the first generation leptons $\nu_e, e$.
It is possible that this is not the case. 
If the scalars $\chi_L, \chi_R$ coupled the first generation
quarks, $u, d$ with the second generation leptons $\nu_\mu, \mu$ then
the decay $N \to \nu_\mu \mu \bar \mu $ would be the dominant decay
mode. Note that the decay rate for this mode might be 
somewhat larger due to the larger $\lambda_2$. The
experimental bound is only slightly weaker,
${\large \tau}_N \stackrel{>}{\sim} 4 \times 10^{31}\ years$ 
at 90\% C.L.\cite{exp}
so the bound on $\lambda_1 \stackrel{\sim}{\lambda}_1 $
is somewhat stronger,
but certainly cannot exclude a symmetry breaking
scale of the order of a TeV.

\section{Gauge interaction mediated rare decays.}

In the alternative $422$ 
model the right handed leptons belong to the
same multiplet as the right handed quarks. This means
that there will be gauge interactions of the form;
\begin{equation}
{\cal{L}}=
{g_s\over\sqrt2}
\bar{D}^{i}_{R}W^{'}_{\mu}\gamma^{\mu}K^{'ij}
l^{j}_{R}\;+\;H.c.,
\end{equation}
where the latin index is a family index 
(so that $D^1_R = d_R, \ell^1_R = e_R \ D^2_R = s_R$ etc, ),
the $W^{'}_{\mu}$ is the coloured electrically charged
2/3 vector gauge boson and $K^{'ij}$ is a C.K.M.
type matrix. In Ref.\cite{foot} 
it was shown that an approximately diagonal $K^{'}$ matrix,
\begin{equation}\left(
\begin{array}{ccc}
1&0&0\\0&1&0\\0&0&1\end{array}\right),
\end{equation}
would lead to $K^0 \to \mu^{\pm} e^{\mp}$ decay faster
than the experimental limit unless $m_{W'} \stackrel{>}{\sim}
140 \ TeV$.
However, as was discussed in Ref.\cite{foot} there
in no compelling reason why $K^{'}$ must be diagonal,
and it was shown that if $K^{'}$ had the approximate form,
\begin{equation}
\left(
\begin{array}{ccc}
1&0&0\\0&0&1\\0&1&0\end{array}\right),
\end{equation}
then the primary constraint on the model is from
$B^0 \to \mu^{\pm} e^{\mp}$ rare decays\footnote{
Such non-standard $K'$ matrices have also
been studied in the context of the usual Pati-Salam
type model, see Ref.\cite{will2}.}. In this 
case the experimental bound on the $SU(4)$ symmetry 
breaking scale is much weaker, $m_{W'} \stackrel{>}{\sim}
1 \ TeV$\cite{foot}. Our purpose now is to examine
all  possible forms for the matrix $K^{'}$ which can
lead to such low symmetry breaking scales. 
As discussed already in the introduction, a TeV symmetry
breaking scale is theoretically suggested to avoid
a gauge hierarchy and also to make the model accessible
to experiments.
Clearly, the rare decays $K_L \to \mu^{\pm} e^{\mp}$ 
must be suppressed sufficiently for a TeV symmetry
breaking scale to occur, and this implies that
the only possible (approximate) forms for $K^{'}$ are;
\begin{eqnarray}&K^{'}_1&=\left(
\begin{array}{ccc}
0&0&1\\Cos\alpha&Sin
\alpha&0\\-Sin\alpha&Cos\alpha&0\end{array}\right)\;\;,\;\;
K^{'}_2=\left(
\begin{array}{ccc}
Cos\beta&Sin\beta&0
\\0&0&1
\\-Sin\beta&Cos\beta&0\end{array}\right)\;\;,\;\;
\nonumber\\
\nonumber\\
&K^{'}_3&=\left(
\begin{array}{ccc}
Cos\gamma&0&Sin\gamma\\-Sin\gamma&0&Cos\gamma
\\0&1&0\end{array}\right)\;\;,\;\;
K^{'}_4=\left(
\begin{array}{ccc}
0&Cos\delta&Sin\delta\\
0&-Sin\delta&Cos\delta\\
1&0&0
\end{array}
\right).
\end{eqnarray}

\noindent
If $K^{'} = K_i^{'}$ ($i=1,...,4$) then the rare decays $K_L \to
\mu^{\pm} e^{\mp}$ are avoided because the $W'$ connects
either the $d$ quark or $s$ quark with the tau lepton. 
However, as we will discuss 
in detail in a moment, in each case rare $B^0$ decays will occur.
The relevant experimental limits are (at 90\% C.L.)\cite{exp},
\begin{eqnarray}
Br({B}_0\rightarrow\tau^\pm e^\mp)
< 5.3\times10^{-4}, \nonumber \\
Br({B}_0\rightarrow\tau^\pm \mu^\mp)
< 8.3\times10^{-4},
\nonumber \\
Br({B}_0\rightarrow\mu^\pm e^\mp)
< 5.9\times10^{-6}.
\label{br}
\end{eqnarray}
We now discuss the four possible cases $K^{'} = K_i^{'}$ in turn:
\vskip 0.5cm
\noindent
1) If the $K^{'} \simeq K^{'}_1$ then
the $B^0 \to \tau^+ \mu^-$ and $B^0 \to \tau^+ e^-$
may occur, which are mediated by the following Feynman diagrams, 

\newcounter{cms}
\setlength{\unitlength}{1mm}
\begin{picture}(80,23)(-20,0)
\put(5,5){\line(1,0){28}}
\put(5,15){\line(1,0){28}}
\put(20,5){\line(0,1){10}}
\put(32,4.1){${\rightarrow}\;\;\mu^-$}
\put(32,14.1){$\rightarrow\;\;\tau^{+}$}
\put(0,4.1){$b$}
\put(0,14.1){$\overline{d}$}
\put(19,8.5){$\uparrow$}
\put(21,8.5){$W^{'+{2\over3}}$}
\put(55,5){\line(1,0){28}}
\put(55,15){\line(1,0){28}}
\put(70,5){\line(0,1){10}}
\put(82,4.1){$\rightarrow\;\;e^{-}$}
\put(82,14.1){$\rightarrow\;\;\tau^{+}$}
\put(50,4.1){$b$}
\put(50,14.1){$\overline{d}$}
\put(69,8.5){$\uparrow$}
\put(71,8.5){$W^{'+{2\over3}}$}
\end{picture}

The decay rate for $B^0\rightarrow\tau^{+}\mu^{-}$,
assuming maximal $\mu$ production for $\alpha=0$,
is calculated from the above Feynman diagram.
This diagram corresponds (after a Fierz rearrangement) 
to the following effective 4 fermion Lagrangian density,
\begin{equation}
{\cal L}^{eff} = {G_X\over\sqrt2}\bar{d}\gamma_{\mu}
(1+\gamma_5)b\bar{\mu}\gamma^{\mu}
(1+\gamma_5)\tau+H.c.,
\end{equation}
where $G_X \equiv \sqrt{2} g_s^2(m_{W'})/8m_{W'}^2$.
From this effective Lagrangian density it is straightforward to
calculate the decay rate, 
\begin{equation}
\Gamma(B^0\rightarrow\tau^{+}\mu^{-}) =
{G_X^2f_B^2\over{8\pi}}m_Bm_{\tau}^2.
\label{zzi}
\end{equation}
Evaluating this using $f_B \approx 150\ MeV$, $m_B \simeq 5.3\
GeV$ and using the measured total decay rate, we find
the branching fraction,
\begin{equation}
Br(B^0 \to \tau^+ \mu^-) \approx
10^{-3} \left({TeV\over{m_{W'}}}\right)^4.
\end{equation}
Thus, from the experimental limits, Eq.(\ref{br}) we
see that $m_{W'} \stackrel{>}{\sim} 1 \ TeV$.
Similar bounds also occur for other values of $\alpha$.
Note that in the case where $\alpha \simeq \pi/2$
the bound comes from the $B^0 \to \tau^+ e^-$ decay.

\vskip 0.5cm
\noindent
2) If the $K^{'} \simeq K^{'}_2$ then
the $B^0 \to e^{\pm} \mu^{\mp}$ decays can
occur via the following Feynman diagrams,

\begin{picture}(80,23)(-20,0)
\put(5,5){\line(1,0){28}}
\put(5,15){\line(1,0){28}}
\put(20,5){\line(0,1){10}}
\put(32,4.1){$\rightarrow\;\;\mu^{-}$}
\put(32,14.1){$\rightarrow\;\;e^{+}$}
\put(0,4.1){$b$}
\put(0,14.1){$\overline{d}$}
\put(19,8.5){$\uparrow$}
\put(21,8.5){$W^{'+{2\over3}}$}
\put(55,5){\line(1,0){28}}
\put(55,15){\line(1,0){28}}
\put(70,5){\line(0,1){10}}
\put(82,4.1){$\rightarrow\;\;e^{-}$}
\put(82,14.1){$\rightarrow\;\;\mu^{+}$}
\put(50,4.1){$b$}
\put(50,14.1){$\overline{d}$}
\put(69,8.5){$\uparrow$}
\put(71,8.5){$W^{'+{2\over3}}$}
\end{picture}

\noindent 
The decay rate for the first process is proportional to
$cos^4{\beta}$ and for the second process it is proportional 
to $sin^4\beta$.  
The Feynman diagrams can easily be evaluated as before,
the only difference is that $m^2_\tau \to m^2_\mu$ in
Eq.(\ref{zzi}). Taking the case $\beta = 0$ then
\begin{eqnarray}
\Gamma(B^0\rightarrow e^{+}\mu^{-}) &=&
{G_X^2f_B^2\over{8\pi}}m_Bm_{\mu}^2, \nonumber \\
\Rightarrow
Br(B^0 \to e^+ \mu^-) &=&
3\times 10^{-6} \left({TeV\over{m_{W'}}}\right)^4.
\label{zzi2}
\end{eqnarray}
Thus, comparing the above rate with
the experimental lower limit, in Eq.(\ref{br})
we see that the limit on the $W'$ mass is also about 1 TeV
in this case (similar bounds also occur for other values 
of $\beta$).

\vskip 0.5cm
\noindent
3) If the $K^{'} \simeq K^{'}_3$ then
the $B^0 \to \mu^- e^+$ and
$B^0 \to \mu^- \tau^+$ decays can occur via
the following Feynman diagrams,

\begin{picture}(80,23)(-20,0)
\put(5,5){\line(1,0){28}}
\put(5,15){\line(1,0){28}}
\put(20,5){\line(0,1){10}}
\put(32,4.1){$\rightarrow\;\;\mu^{-}$}
\put(32,14.1){${\rightarrow}e^{+}$}
\put(0,4.1){$b$}
\put(0,14.1){$\overline{d}$}
\put(19,8.5){$\uparrow$}
\put(21,8.5){$W^{'+{2\over3}}$}
\put(55,5){\line(1,0){28}}
\put(55,15){\line(1,0){28}}
\put(70,5){\line(0,1){10}}
\put(82,4.1){$\rightarrow\;\mu^{-}$}
\put(82,14.1){$\rightarrow\;\tau^{+}$}
\put(50,4.1){$b$}
\put(50,14.1){$\overline{d}$}
\put(69,8.5){$\uparrow$}
\put(71,8.5){$W^{'+{2\over3}}$}
\end{picture}

\noindent 
The rate for the first process is proportional to
$cos^2 \gamma$ and for the second process it is
proportional to $sin^2\gamma$.
These processes  
are similar to cases already discussed, and
the lower bound in this case is therefore also about 1 TeV.

\vskip 0.5cm
\noindent
4) If the $K^{'} \simeq K^{'}_4$ then
the $B^0 \to \mu^+ e^-$ and
$B^0 \to \tau^+ e^-$ decays can occur, and
the bound from these decays, being similar
to processes already discussed is also about a TeV.
However, in this case there is another possible rare decay
which is $K_L \rightarrow\mu^+\mu^-$.
This decay rate is proportional 
to the factor $sin^2{\delta}cos^2\delta$,
\begin{eqnarray}
\Gamma(K_L \rightarrow\mu^{\pm}\mu^{\mp}) &=&
sin^2{\delta}cos^2\delta{G_X^2f_K^2\over{4\pi}}m_Km_{\mu}^2,
\nonumber \\
\Rightarrow Br(K_L \to \mu^{\pm} \mu^{\mp})
&\approx & 5\times 10^{-3}
\left({TeV\over{m_{W'}}}\right)^4 \ for \ \delta = {\pi \over 4}.
\label{sd}
\end{eqnarray}
The measured branching ratio is\cite{exp};
\begin{equation}
Br(K_L \rightarrow\mu^+\mu^-)
= (7.2\pm 0.5) \times 10^{-9}.
\end{equation}
Conservatively, demanding that the $W'$ contribution Eq.(\ref{sd})
to be less than the total branching fraction, implies the limit,
$m_{W'} \stackrel{>}{\sim} 30\ TeV$, for the maximal
case where $\delta = \pi/4$.

We briefly summarise the main results in the following table;

{\begin{center}
\begin{tabular}{|l|l|l|}
\hline
{\em Matrix}$\;\;\;\;\;\;\;\;\;\;\;$
&{\em Process}$\;\;\;\;\;\;\;\;\;\;\;\;$
&{\em Bound}$\;\;\;\;\;\;\;\;\;\;\;$ \\
\hline
$K_1^{'}\;;\;\alpha=0$&$B^0{\rightarrow}\mu^{-}\tau^{+}$
&$W^{'}{\stackrel{>}{\sim}}1$TeV\\
\hline
$K_1^{'}\;;\;\alpha={\pi\over 2}$&$B^0{\rightarrow} e^{-}\tau^{+}$
&$W^{'}{\stackrel{>}{\sim}}1$TeV\\
\hline
$K_2^{'}\;\;;\beta=0$&$B^0{\rightarrow}e^{+}\mu^{-}$
&$W^{'}{\stackrel{>}{\sim}}1$TeV\\
\hline
$K_2^{'}\;\;;\beta={\pi\over2}$&$B^0{\rightarrow} e^{-}\mu^{+}$
&$W^{'}{\stackrel{>}{\sim}}1$TeV\\
\hline
$K_3^{'}\;\;;\gamma=0$&$B^0{\rightarrow}e^{+}\mu^{-}$
&$W^{'}{\stackrel{>}{\sim}}1$TeV\\
\hline
$K_3^{'}\;\;;\gamma={\pi\over2}$&$B^0{\rightarrow}\tau^{+}\mu^{-}$
&$W^{'}{\stackrel{>}{\sim}}1$TeV\\
\hline
$K_4^{'}\;\;;\delta=0$&$B^0{\rightarrow}\mu^{+}e^{-}$
&$W^{'}{\stackrel{>}{\sim}}1$TeV\\
\hline
$K_4^{'}\;\;;\delta={\pi\over2}$&$B^0{\rightarrow}\tau^{+}e^{-}$
&$W^{'}{\stackrel{>}{\sim}}1$TeV\\
\hline
$K_4^{'}\;\;;\delta={\pi\over4}$&$K^0{\rightarrow}
\mu^{+}\mu^{-}$
&$W^{'}{\stackrel{>}{\sim}}30$TeV\\
\hline
\end{tabular}\end{center}}

\vskip 0.5cm
\section{Naturally small neutrino masses}
\vskip 0.5cm

In the $422$ model there are four electrically neutral Weyl states
per generation, $\nu_L, \nu_R, E_{L,R}^0$.
Thus the masses for the neutral leptons will
be described by a $12 \times 12$ mass matrix.
The $E_{L,R}^0$ gain masses from the large VEV $w_R$ and
are expected to be quite heavy (experimentally we
know that they must be heavier than about $m_Z/2 \approx 45\ GeV$).
While the approximately sterile (i.e.
$SU(2)_L \otimes U(1)_Y$ singlet) $\nu_R$ states gain
masses by mixing with the $E$ leptons (see below for more details).
At tree level the ordinary neutrinos (i.e. the $\nu_L$ states)
are massless. This is quite easy to see,
because the masses of the fermions arise from the Lagrangian
density Eq.(\ref{4}), and the $\nu_L$ states do not couple
to any VEV.

In order to gain insight into the neutrino masses, let
us first consider the toy case of just one generation,
with just the usual first generation states (together with
the exotic $E$ leptons).  In this case
the tree level neutrino mass matrix, which can
be obtained from Eq.(\ref{4}), has the form:
\begin{equation}
{\cal L}_{tree} = \bar \psi_L M (\psi_L)^c + H.c.,
\end{equation}
where
\begin{equation}
\psi^T_L = (\nu_L, (\nu_R)^c, E_L^0, (E_R^0)^c),
\end{equation}
and
\begin{equation}
M = \left(
\begin{array}{cccc}
0&0&0&0\\
0&0&m_u&m_e\\
0&m_u&0&m_E\\
0&m_e&m_E&0
\end{array}\right).
\end{equation}
Thus, at the tree level the ordinary neutrino
$\nu_L$ is massless. The 'light' sterile $\nu_R$ state
has mass
\begin{equation}
m_{\nu_R} \simeq {2m_u m_e \over m_E}.
\end{equation}
At one loop, there are important corrections to this
mass matrix. In Ref.\cite{foot} only one such 
correction ($m_M$) was considered. Here we do a 
better job by including all possible 1-loop (gauge) corrections
involving $\nu_L$. In particular, the mass terms
\begin{equation}
{\cal L}^{eff}_{1-loop} = 
m_M\bar \nu_L (\nu_L)^c + m_D\bar \nu_L \nu_R 
+ m_{\nu E}\bar \nu_L (E_L^0)^c + H.c.,
\end{equation}
are generated from the Feynman diagrams, Fig 2,3,4.
Evaluating these diagrams,
\begin{eqnarray}
m_M &=& m_e m_d m_E {g_R g_L \over 8\pi^2}\left[
{\mu^2 \over m^2_{W_R}}\right]
\left[ {log {m^2_{W_R} \over m_E^2} \over
m^2_{W_R} - m^2_{E}}
- {log {m^2_{W_L} \over m_E^2} \over
m^2_{W_L} - m^2_{E}}
\right], \nonumber \\
m_D &=& m_e{g_R g_L \over 8\pi^2}\left[
{\mu^2 \over m^2_{W_R}}\right]
log\left({m_{W_R}^2 \over m_{W_L}^2}\right), \nonumber \\ 
m_{\nu E} &=& 
m_E {g_R g_L \over 8\pi^2}\left[
{\mu^2 \over m^2_{W_R}}\right]
\left[ log\left({m_{W_R}^2 \over m_{W_L}^2}\right)
+ {m_E^2 
log\left( {m_{W_L}^2 \over m_E^2}\right) 
\over m_E^2 - m_{W_L}^2} - {m_E^2 
log\left( {m_{W_R}^2 \over m_E^2}\right) 
\over m_E^2 - m_{W_R}^2} 
\right],
\end{eqnarray}
where $\mu^2 \equiv g_R g_L u_1 u_2$ is the $W_L - W_R$ mixing mass.
Including these radiatively generated mass terms, the
effective mass matrix becomes
\begin{equation}
M = \left(
\begin{array}{cccc}
m_M&m_D&m_{\nu E}&0\\
m_D&0&m_u&m_e\\
m_{\nu E}&m_u&0&m_E\\
0&m_e&m_E&0
\end{array}\right).
\label{yyt}
\end{equation}
The effect of this is to give the neutrino $\nu_L$ a 
small Majorana mass, given approximately by 
\begin{equation}
m_{\nu} \simeq {Det(M)\over 2m_e m_u m_E}, 
\end{equation}
that is,
\begin{equation}
m_{\nu} \simeq m_M + {m_D^2m_E\over 2m_em_u}
+ {m^2_{\nu E} m_e \over 2m_u m_E}
- {m_{\nu E} m_D \over m_u}.
\end{equation}
Actually no precise predictions can be made for the neutrino
masses, due, for example, to the unknown masses of the heavy $E_i^0$
leptons.  Nevertheless it is possible to show that the 
neutrino masses are naturally light. From Eq.(\ref{4})
the VEV's $u_1, u_2$ can be related to the bottom and top 
quark masses as follows 
\begin{equation}
m_b = \lambda_3 u_1 + \lambda_4 u_2, 
\ m_t = \lambda_3 u_2 + \lambda_4 u_1.
\end{equation}
It follows that
\begin{equation}
{u_1 u_2 \over u_1^2 + u_2^2} \sim {m_b \over m_t}.
\end{equation}
Hence
\begin{equation}
{\mu^2 \over m_{W_R}^2} \sim 
{1 \over 2\sqrt{3}}{m_{W_L}^2 \over m_{W_R}^2}{m_b \over m_t},
\end{equation}
where we have used $g_R \simeq g_L/\sqrt{3}$\cite{foot} and
$m_{W_L}^2 = {1 \over 2}g_L^2(u_1^2 + u_2^2 + w^2_L) \sim 
{1 \over 2}g_L^2(u_1^2 + u_2^2)$.
Thus, we have
\begin{eqnarray}
m_M &\sim &{m_e m_d \over m_E}{g_L^2 \over (4\pi)^2}
{m_{W_L}^2 \over m_{W_R}^2}{m_b \over m_t},
\nonumber \\
m_D &\sim& {m_E \over m_d}m_M,
\nonumber \\
m_{\nu E} &\sim& {m_E \over m_e}m_D.
\end{eqnarray}
Hitherto we have studied only the one generation case. 
Of course the full neutral lepton mass matrix will be a 
$12 \times 12$ generalisation of Eq.(\ref{yyt}).
While the general mass matrix is obviously quite
complicated, with many free parameters, it is
still possible to place an upper limit on the largest
possible (ordinary) neutrino mass.
This will occur when $m_e \to m_\tau$
and $m_d \to m_b$ (with $m_u \to m_u$, unchanged). 
In this case
\begin{equation}
\left. m_M \right|_{max} \sim 
 {m_\tau m_b \over m_E}{g_L^2 \over (4\pi)^2}
{m_{W_L}^2 \over m_{W_R}^2}{m_b \over m_t}
\sim 50 \left( {TeV \over m_{W_R} }\right)^2
\left({100 GeV \over m_E}\right) \ eV,
\end{equation}
and
\begin{eqnarray}
\left. {m_D^2 m_E \over m_\tau m_u}\right|_{max} &\sim & 
\left. {m_{\nu E}^2 m_\tau \over m_E m_u}\right|_{max} 
\sim \left. {m_{\nu E} m_D \over  m_u}\right|_{max}, 
\nonumber \\
& \sim &
m_\tau \left( {g_L^4 \over (4\pi)^4}\right) \left({m^4_{W_L}
\over m^4_{W_R}}\right) {m_b^2 \over m^2_t}{m_E \over m_u}
\sim 20 \left( {TeV \over m_{W_R}}\right)^4 \left({m_E \over 100 GeV}
\right)
\ eV.
\end{eqnarray}
Thus the upper limit on the neutrino mass is naturally 
light (i.e. less than about $50 \ eV$)
despite the low TeV symmetry breaking scale of the model.
Of course all three neutrinos may be considerably ligher
than this maximum mass, such information will depend on the
parameters of the full $12 \times 12$
neutral lepton mass matrix.  Finally note
that in addition to three light neutrinos, the model has three 
heavier sterile neutrinos the $\nu_R$'s, and the heavy leptons
$E_i^0$.

\vskip 0.5cm
\section{Conclusion}
\vskip 0.5cm
We have studied the alternative $SU(4)\otimes SU(2)_L
\otimes SU(2)_R$ gauge model which allows unification of
the quarks and leptons at the TeV scale.
We have shown that the leading nucleon decay mode
in this model is (bound) neutron decay,
$N \to \nu \ell \bar \ell$ (where
$\ell = e, \mu$). While current experimental bounds
on bound neutron decay
cannot exclude a TeV symmetry breaking scale, such
experimental searches can potentially test the model.
More important tests are expected to come from the up-coming
B factory experiments. From improved limits (or discoveries!) of rare
B decays, such as $B^0 \to e^{\pm} \mu^{\mp}$,
$B^0 \to e^{\pm} \tau^{\mp}$ and $B^0 \to \mu^{\pm} \tau^{\mp}$,
much of the most interesting region of parameter space
where the $SU(4)$ symmetry breaking scale
is in the TeV range will be covered.
Finally, the neutrino masses are radiatively generated and are
naturally quite light, with an upper limit of about $50$ eV.

\vskip 1cm
\noindent
{\bf Acknowledgements}
\vskip 0.5cm
R.F. is an Australian Research Fellow.

\newpage
\vskip 1cm
\noindent
{\bf Figure Captions}
\vskip 0.5cm
\noindent
Figure 1: Feynman diagram for the scalar mediated neutron
decay $N \to e^+ e^- \nu_e$.

\vskip 0.5cm
\noindent
Figure 2: 1-loop Feynman diagram which leads to small  
neutrino Majorana mass term. 
(The $W_L W_R$ mixing mass squared is obtained from
${\cal L} = (D_{\mu} \langle \phi \rangle )^{\dagger} 
D^{\mu} \langle \phi \rangle$ and is given 
by $\mu^2 = g_R g_L u_1 u_2$).

\vskip 0.5cm
\noindent
Figure 3: 1-loop Feynman diagram leading
to the mass term $\bar \nu_L \nu_R$.

\vskip 0.5cm
\noindent
Figure 4: 1-loop Feynman diagram leading to
the neutrino mixing term $\nu_L (E_L^0)^c$.

\newpage
{\bf Figure 1}
\begin{picture}(130,80)(-40,-20)
\put(0,5){\line(1,0){30}}
\put(10,7){$u_b$}
\put(5,16){\line(1,0){30}}
\put(10,18){$d_g$}
\put(0,27){\line(1,0){20}}
\put(10,29){$d_r$}
\put(20,27){\line(4,1){46}}
\put(19,29){$\lambda_2$}
\multiput(20,27)(8,0){6}{\line(1,0){4}}
\multiput(20,16)(8,2){6}{\line(4,1){4}}
\put(19,12){$\lambda_2$}
\put(41,0){$\lambda_1$}
\multiput(42,5)(6,6){4}{\line(1,1){4}}
\put(30,16){\line(1,0){40}}
\put(30,5){\line(1,0){40}}
\put(67,4.1){$\rightarrow\;\;\;e^+$}
\put(67,15.1){$\rightarrow\;\;\;e^-$}
\put(73,38){${\nu_e}$}
\put(48,29){$\chi_{L}^{-{1\over2}}$}
\put(26,20){$\chi_{L}^{+{1\over2}}$}
\put(38,8){$\chi_{R}^{-{1\over2}}$}
\put(65.5,37.6){$\rightarrow$}
\put(0,16){\oval(10,22)}
\put(-10,16){\line(1,0){10}}
\put(-14,15){$N$}
\put(-3,15){$\rightarrow$}
\end{picture}

\newpage
\epsfig{file=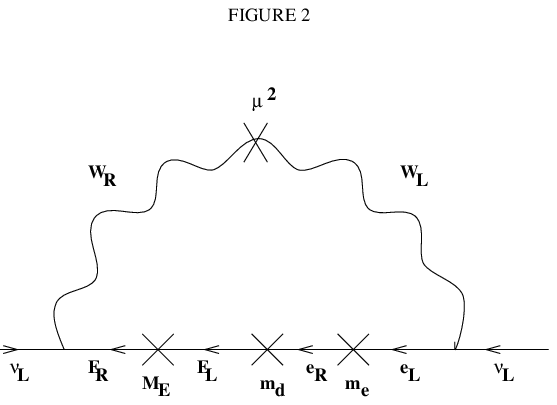,width=15cm}
\newpage
\epsfig{file=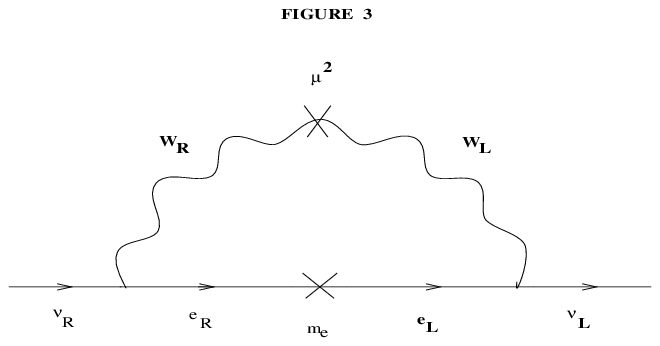,width=15cm}
\newpage
\epsfig{file=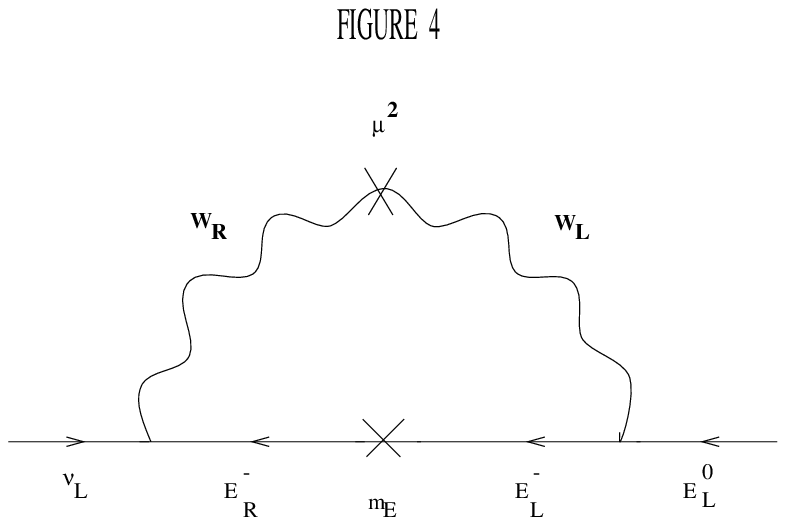,width=15cm}
\end{document}